# Addressing the Recruitment and Retention of Female Students in Computer Science at Third Level

Dr. Susan McKeever, Dr Deirdre Lillis

*School of Computer Sciemce, Technological University Dublin*

**Abstract**

In the School of Computing at the Dublin Institute of Technology[ii] (DIT), Ireland, we undertook our "Computer Science for All" (CS4All) initiative, a five year strategy to implement structural reforms at Faculty level, to address recruitment and retention issues of female undergraduate computer science (CS) students. Since 2012, under CS4All we implemented a variety of reforms to improve student retention, set up a new CS programme to attract more female students, and delivered changes to promote a sense of community amongst our female students. We have made significant improvements. For example, we have achieved a dramatic improvement in retention rising from 45% to 89% in first year progression rates. Our new hybrid CS International programme has more than double the percentage of females first year enrolments in comparison to our other undergraduate programmes. As at 2018, we continue to roll out the remaining parts of CS4All within our School.

## 1. Introduction

The CS4All initiative emerged from a previous review of our school in 2011. At that time, we faced the issue of decreasing numbers of females taking our undergraduate programmes, hand-in-hand with poor student retention - problems common across the third level CS sector in Ireland and in many countries internationally. In 2011, our progression rate from first year to second year on our undergraduate degrees was averaging 45%, and our flagship BSc in Computer Science programme had just 10% females enrolment to first year. Rather than continue to rely on individual staff driven initiatives, as a School we took the decision to address gender balance and retention with School-driven strategic structural reforms. From this, the CS4All initiative emerged - with a multi-purpose agenda: How do we increase the numbers of female students coming to our CS undergraduate programmes? How do we improve the experience for these students, and reduce the numbers failing to progress in the critical first year? How do we spread the message that computer science is a good choice for females, as well as males? Our CS4All programme aimed to address this problem, split into several phases: (1) Analysis and Actions (2) Quick Win actions (3) Strategic actions, as explained next.

**CS4ALL Initiative Description**

**Phase 1:** *Analysis and Actions:* To avoid a scatter-gun approach of disparate actions, we used an evidence-based approach to identify both actions with an immediate benefit and longer term systematic strategic actions. This approach was structured as follows:

(1) *Retention*: We analysed own retention situation to do a 'root-cause analysis' of why students fail to progress on the programme. Our first-year progression rates from 2004/05 to 2008/09 years ranged between 30% and 50%, averaging 45%. We identified that students who progressed at the first attempt in first year had a far higher chance of finishing their degree. From staff consultations, we noted the critical link between attendance and successful progression.  We also found a strong link between programming skills in first year and success in later year. From this, we identified the actions that CS4ALL focused on: Critical skills; Attendance monitoring;  Incentives to participate in practical work; Assessment / grades turning; Building a culture of attendance and participation.

(2) *Recruitment*:  Our starting point was to understand which ICT programmes were attracting the highest proportions of females across Ireland. We obtained a "gender balance enrolment to ICT courses" dataset from the Higher Education Authority data (also available at [1])  to analyse the patterns of female enrolment on ICT programmes across Ireland. The data showed than 40% of computing programmes had a female intake of less than 10% (see **Figure 1**). .Analysing the programmes that had higher female enrolments, we noted that traditional Computer Science programmes were attracting the lowest proportion of females.  The programmes attracting highest rates were those offering an additional domain - which we term "hybrid" programmes - such as "Computer Science and Languages" in Trinity College Dublin with 33% female enrolments.  To illustrate this, the word cloud shown in **Figure 2** shows the relative popularity of word occurrences in computing programme titles of programmes with high female enrolment ((> 20% females in year 1).  Note the high profile of non-computing words such as "marketing", "media", "management".  Our findings on female attraction to hybrid programmes ties in with research literature that suggests that female students prefer to use a wider range of their skills, avoiding the pure specialisations of science [2].  Our actions in this domain were to set up a new undergraduate programme to tap into the attraction of females to hybrid prograrmmes.

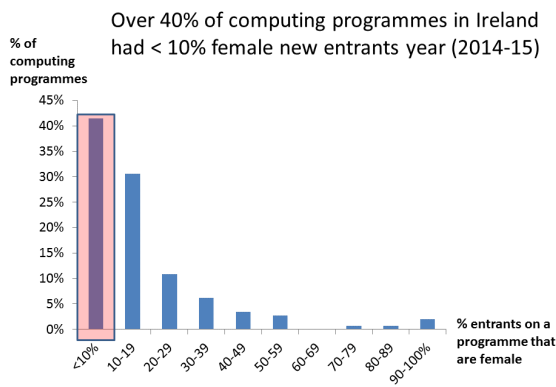

**Figure 1:** % of females entering Year 1 of ICT third level programmes (levels 6-8) in Ireland in 2014-15, based on Higher Education Authority data [1]

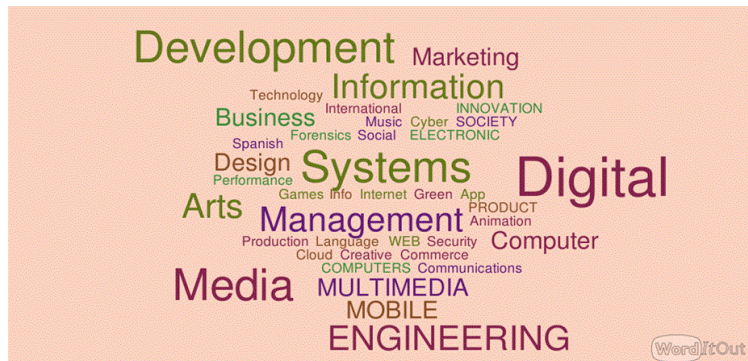

**Figure 2:** A word cloud showing the relative occurrences of words in the titles of level 8 ICT programmes in Ireland in 2014-15. Only programmes with a high (>20% female enrolment) are shown. Derived from Higher Education Authority data [1]

(3) *The Female Student Experience*: we met with and surveyed our own female student cohort via arranged focus groups. Having never done this before, we wanted to find out how whether their college experience was being impacted (either positively or negatively) by being in the minority gender on their programmes. Key feedback from the group was that many females did not want 'special treatment' on the basis of their minority status. This informed our CS4All approach. We identified that some of our students felt isolated and found it difficult to find a peer group. We also noted that many of our female students had "discovered" computer science relatively randomly - via a talk in school, or participation at a promotional event. Our actions from these consultations were to set up a female community for our School of Computer Science students, examine how to reduce isolation for students and to find opportunities to promote Computer Science to non-tech females or younger females.

**Phase 2: Quick *Win Implementation*:** The purpose of this phase was to implement the actions that could be achieved immediately - based on the actions identified from the Analysis and Actions Identification phase.

To tackle *Retention*, we introduced:
- Random first year attendance monitoring, with rapid active follow-up by the mentors with each student;
- Weekly grading of practical lab work every week to encourage participation and boost practical skills;
- Written exam thresholds on all programming modules to ensure solid theoretical foundation;
- Incorporation of the key message of participation and attendance into first year induction;
- Revised module sequences to balance practical vs theory modules evenly across semesters in year 1.

To boost *Recruitment*, we:
- Established a new hybrid programme, the BSc Computer Science (International), where each student studies a language (Chinese, German, Korean) as part of their study. This programme has a strong international component, with an overseas work placement or Erasmus component.

- Set up free coding courses for non-tech female students in DIT, in co-operation with the CodeFirstGirls social enterprise organisation [3]; We are the first third level in Ireland to roll out CodeFirstGirls courses.
- Run the annual "Computing Academy" for transition year students, with 50/50 gender balance on attendance [5];
- Reviewed our traditional BSc. in Computer Science programme to identify areas, such as international work and study placements, to make it more attractive to females.

*To improve the Female Student experience, we introduced:*
- A practice that each female student will be placed in lab groups with at least one other female as this was a particular pinch point of isolation for female students on their own;
- Introduced undergraduate student lab assistants into first year programming labs, in order to overcome the "confidence barrier" in asking questions.
- The creation of a female student community through regular hosted meet-ups organised by the School

**Phase 3: Strategic Actions**  This step focused on the actions that have a longer term strategic return:
- We embedded gender as a principal consideration into- School Planning & Review processes, school meetings, promotion opportunities and personal development planning sessions with staff (raising awareness amongst all staff of the importance of this initiative)
- We provided school-level resourcing, factored initiatives into staff timetables, budgets and planning to ensure this was no longer an ad-hoc activity.
- We appointed a CS4ALL Champion who spearheaded the programme of reform.
- We coordinated and lead the successful Erasmus+ Knowledge Alliance project Hublinked with 13 international industry and academic partners which includes gender as one of four key objectives (European funding of €1,000,000 from 2017 to 2020);
- We established the Third Level Computing Action group, an alliance of Schools of Computer Science throughout the third level sector in Ireland, in order to promote sharing of best practice through the sector, rather than each school working in isolation.
- We continue with our strategic plan for gender balance, as presented at our recent School Review.

**2. Evidence of Impact of the CS4All Initiative**

One of the difficulties that we encountered prior to CS4All was the difficulty of measuring impact of disparate initiatives. For example, after running a transition year event for female students, we had no way of measuring the direct impact of such an initiative on recruitment. This is a similar difficulty faced by other third levels, as highlighted by the Third Level Computing Action Group that we have established as part of CS4All.

**(i) Female student recruitment**
*New Hybrid Bachelors programme and Erasmus implementation*
Our new BSc in Computer Science (International) is now in its fourth year of operation. The proportion of females on the programme averages at 22% over the four years since it was established. This is more than double the female enrolment rate on our BSc Computer Science programme. Our new programme has led to a Double Degree arrangements with four international partners in Germany, Finland and Korea.

An emerging trend is the attraction of Erasmus study for female students. Firstly, the higher female enrolment rates on to our new BSc in Computer Science (International) of which Erasmus is a key component is an indicator of this. Having seen this trend, we added an Erasmus option to our traditional BSc Computer Science. In the past two years, the proportion of Erasmus students on our traditional BSc who are female has been 25% (201617) and 28% (201718) respectively. This is an early indicator but we will build on this trend if we see that Erasmus may be a carrot for improved female recruitment.

*Free Coding courses for female non tech DIT students*

As explained in our letter of support from CodeFirstGirls, we have hosted four semesters of CodeFirstGirls [3] coding courses, with more than 120 girls having completed a HTML or python programme with our School [4]. These courses are open to any DIT female student who is not on a computing programme. We have reached application rates of 8 applications to 1 place, demonstrating a real appetite amongst female students throughout DIT to try out technology and coding. We have had wide spread of participation throughout DIT from arts/ humanities, business, science and engineering. In our last two semesters of feedback metrics, 100% of females on the courses said they would consider a career in technology and 83% found the coding course useful. Our fifth set of courses starts in September 2018.

*Computing Academy, CanSat and other transition year events*

We secured national funding to run a highly successful gender balance Computing Academy for second level students (Aged 15-17) which has consistently had a high demand for places. This annual Computing Academy week, run in May each year, is a regular diary event for the School, run by lecturing staff. We have hosted more than 300 female students over the years. The feedback from this event is highly positive with schools contacting us each year to attempt to get places for their students and provides a great opportunity for all staff in the School to become involved in the CS4All work plan.

**(ii) Retention**

Our first year progression rates from 2004/05 to 2008/09 years ranged between 30% and 50%, averaging at 45%. Based on manual analysis of class level data, we could see that retention for male and females was on the same levels, with the problem existing across the full student base.

We have made dramatic improvements in the past five years in our progression rates, particularly for first years. At an average 89% progression from year 1 to year 2, we now have the highest progression rates for Computer Science in the higher education sector in Ireland. **Figure 4** shows the year on year retention rate improvement on our main BSc Computer Science programme over the past 5 years. We now have access to gender specific reports, allowing us to systematically examine gender breakdown back to 201415. As shown in **Figure 5**, female retention rates are slightly higher than males, based on looking at the past three years. This is encouraging but we need to keep focused on reducing retention for all students.

At our recent five yearly School Review, we were commended as an exemplar for the Institute for our proactive approach to improving retention and the substantial gains we have made. A key lesson learning from our retention efforts is that CS4All is not just about gender, it is about an inclusive strategy where special attention is paid to gender.

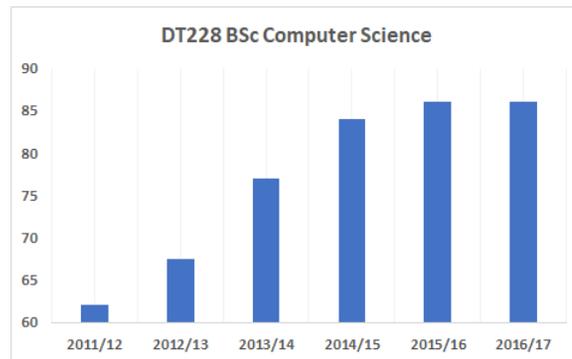

**Figure 4:** Retention rate (% students in year 1 progressing to year 2 in same academic year) by year.

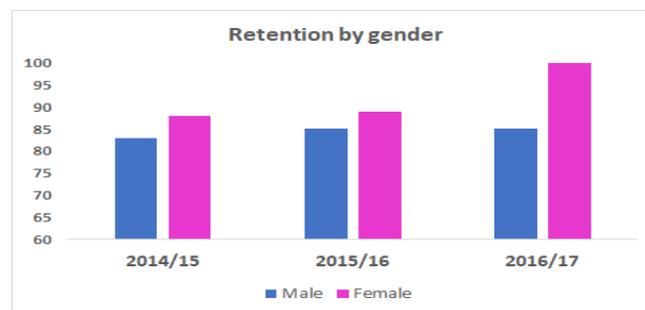

**Figure 5:** Retention rate ((% students in year 1 progressing to year 2 in same academic year) by year by gender BSc Computer Science

**(iii) Female student experience**
At this point, we have an active female undergraduate community within our school. We have had significant female achievements in the School, including the winning of the Global Undergraduate Award (national winner) and a disproportionately high representation of final year female prize-winners at our annual Project Fair, judged by external judges each year.

**Future Actions**
Addressing gender balance is now firmly embedded as part of our School strategy. We are looking forward to building further on our gender balance initiatives to date. Initiatives planned for 2018 onwards include:

We have been selected as an exemplar by our Institute for first-wave Athena Swan accreditation. This is one of the leading international accreditations which recognises the commitment of Higher Education Institutions to advancing the careers of women in science, technology, engineering, mathematics and medicine (STEMM) employment in higher education and research. (target date 2019).

Our eSTeEM initiative [6] for our female School of Computing students starts in September 2018.  This is a scheme to support industry mentorship and provision of role models of all our female CS students.

We will continue with our establishment of the Third Level Computing group on gender balance. We have gathered representatives from 17 universities/ IoTs across Ireland. The purpose of the group is to share best practice on student gender balance initiatives in a "learn from each other" approach. The idea sharing stage has been completed. The first cross-third level meeting is scheduled for June 20th 2018.

---

[i] https://www.informatics-europe.org/awards/40-awards/minerva-informatics-equality-award/winners/503-minerva-winner-2019.html

[ii] Technological University Dublin (TU Dublin) was established on 1st January 2019, arising from a merger of Dublin Institute of Technology, Institute of Technology Tallaght and the Institute of Technology Blanchardstown. DIT was a member of Informatics Europe.